\documentclass[%
 reprint,
 amsmath,amssymb,
 aps,
]{revtex4-2}

\usepackage{graphicx}% Include figure files
\usepackage{caption}
\usepackage{subcaption}
\usepackage{float}
\usepackage{dcolumn}% Align table columns on decimal point
\usepackage{bm}% bold math
\begin{document}

\preprint{APS/123-QED}

\title{Convolutional method for data assimilation\\ An improved method on 
neuronal electrophysiological data  }% Force line breaks with \\
% \thanks{A footnote to the article title}%

\author{Dawei Li}
%  \altaffiliation[Also at ]{Physics Department, XYZ University.}%Lines break automatically or can be forced with \\

%  \email{Second.Author@institution.edu}
\affiliation{%
 Department of Physics, University of California, San Diego
}%

% \collaboration{MUSO Collaboration}%\noaffiliation

\author{Henry D. I. Abarbanel}
%  \homepage{http://www.Second.institution.edu/~Charlie.Author}
\affiliation{
 Department of Physics, University of California, San Diego
}%
\affiliation{
 Marine Physical Laboratory (Scripps Institution of Oceanography), University of California, San Diego 
}%
\affiliation{
 Center for Engineered Natural Intelligence, University of California, San Diego
}%
% \author{Delta Author}
% \affiliation{%
%  Authors' institution and/or address\\
%  This line break forced with \textbackslash\textbackslash
% }%

% \collaboration{CLEO Collaboration}%\noaffiliation

\date{\today}% It is always \today, today,
             %  but any date may be explicitly specified

\begin{abstract}
 
% \begin{description}
% \item[Usage]
% Secondary publications and information retrieval purposes.
% \item[Structure]
% You may use the \texttt{description} environment to structure your abstract;
% use the optional argument of the \verb+\item+ command to give the category of each item. 
% \end{description}
\end{abstract}

%\keywords{Suggested keywords}%Use showkeys class option if keyword
                              %display desired
\maketitle

%\tableofcontents

\section{\label{sec:level1}Introduction}
Data Assimilation (DA) is a method that transfers information from observed data of a dynamical system to a nonlinear dynamical model proposed for the system\cite{abarbanel2022statistical}. With the measurements and a model structure with unknown parameters, we can estimate model parameters and dynamical state variables by synchronizing the estimation with the data(optimization), predicting the future behavior of the nonlinear system. Typical examples of applications include weather forecasts, oceanography, and electrophysiology. Traditionally, the cost function of optimization is based on root-mean-square(RMS) that compares the magnitude of the values at each point along the time series. However, this magnitude-based design can not evaluate all performance quality since timing-based information is vital for some systems, especially in electrophysiology. 

Neuron activity contains action potentials(spikes) and subthreshold behaviors, from which spike timing is believed to encode essential data. The precise spike timing guarantees the transmission quality of temporal and spatial information\cite{panzeri2001role}. The relative spike timing between the presynaptic and postsynaptic neurons contributes to synaptic plasticity\cite{bi1998synaptic}. When we perform data assimilation on electrophysiology data, slight mismatches of sharp spikes usually appear due to an imperfect model. The extent of such a mismatch also determined the quality of estimation and prediction, which suggests the necessity of improving the traditional cost function with the ability to measure the timing-based correlation besides magnitude-based comparison. 

To measure the timing-based correlation, several measures of spike train synchrony were proposed in previous works. Victor and Purpura \cite{victor1997metric} proposed a method that involves representing each spike train as a point in a metric space, where the distance between two points is a measure of the similarity between the corresponding spike trains. The authors use a topological concept known as the ``edit distance" to calculate the similarity between two spike trains. Taking the timing and ordering of spikes into account, the edit distance measures the minimum number of operations required to transform one spike train into the other, where each operation can be a deletion, insertion, or substitution of a spike. Based on Victor and Purpura's work, Kreuz\cite{kreuz2013monitoring}proposes a new method called ``SPIKE-distance" for measuring the degree of synchrony between spike trains. The method considers temporal precision by calculating the cross-spike intervals that reflect the degree of synchrony. The SPIKE-distance method is more robust, computationally efficient, and easier to interpret than the Victor-Purpura distance method, and it has become widely used in the field of neuroscience for studying neural processing and communication. However, obtaining the spike distance makes the measure of synchrony nonanalytic, which is not applicable for Gradient-based optimization methods. Using convolution kernel would solve this issue.

In 1998, Szucs\cite{szHucs1998applications} introduced the concept of SDF, a continuous function that represents the firing pattern of a neuron by convolving a Gaussian kernel function with the spike train. This transformation from a discrete spike train to a continuous function provides a more detailed and informative representation of the temporal dynamics of neuronal firing. SDF has since become a widely used method for analyzing and visualizing neural activity. Later, in 2001, Van Rossum\cite{van2001novel} applied a similar idea to measure the synchrony of spike trains. By transforming the spike trains into continuous functions using an exponentially decaying kernel function, he was able to calculate a distance measure between the spike trains, called ``van Rossum distance". This method provides a measure of the degree of synchrony between two or more neurons based on their spike timings. van Rossum distance is a useful tool for analyzing the functional connectivity of neural networks and has been used in a variety of studies to investigate the properties of neural circuits.

Spike trains are discrete, binary signals that represent the timing of individual neural spikes, while the neural electrophysiology data measured by electrodes can be continuous, analog signals that represent the voltage fluctuations in a neuron over time. In Szucs and van Rossum's works, when applying a convolution kernel on spike trains, the kernel is typically designed to capture only the temporal structure of the neural spike events, and the resulting convolution output will also be a spike train, with each spike representing the result of the kernel operation at a particular time point. The measure of synchrony only describes the degree of synchrony of the temporal structures of the neural spike events. Unlike the previous works, by applying a convolution kernel on continuous signal with detailed spike and subthreshold form, we can synchronize not only the temporal structure of the neural spike events but also the detailed patterns of the neural activities. This measure of synchrony that combines timing and magnitude satisfied our data assimilation purpose and our data assimilation method will help refine the dynamical model and give predictions using this new measure of synchrony.

In this article, we developed an improved convolutional method for data assimilation that balanced the magnitude-based and timing-based measures. A Gaussian filter has been added to the data and its estimate, making the sharp spikes wider. The spike simultaneousness can be inferred from the level of overlap. The advantage of the new method is verified by applying it to twin experiments and experimental data, and comparing the results with the traditional method. We also tested its benefit in evaluating predictions. We added further guidance on selecting the hyperparameter and discussed the possible annealing improvement.

\section{Methods}
\subsection{Traditional Method}
In a $D$-dimensional dynamical model, we have $D$ state variables $x_d(t)$; $d = 1,2,...,D$. $\bm x(t) = \{x_1(t), x_2(t), ...,x_D(t)\}$, among which $L$ state variables is observed. The corresponding observations(data) $y_l(t)$; $l = 1,2,...,L$. $\bm y(t) = \{y_1(t), y_2(t), ...,y_L(t)\}$ are measured at discrete time points $t = \{t_0,t_1,...,t_N \}, t_n=t_0+n\Delta t; n=0,1,...,N$. The time-independent unknown parameters of the model are defined as $\bm p = \{p_1,p_2, . . . ,p_{N_p} \}$.
$D \geqslant L$ and typically $D \gg L$.
The model is moved along in time by step size $\Delta t$ through the dynamical equations:
\begin{equation*}
\dfrac{dx_d(t)}{dt}=F_d(\bm x(t),\bm p); d=1,2,...,D
\end{equation*}
\begin{equation*}
x_d(n+1)=f_d(\bm x(n),\bm p)=x_d(n) + \Delta t (F_d(\bm x(n),\bm p)
\end{equation*}
We call the full collection of state variables along the time series the path vector $\bm X = \{\bm x(t_0),\bm x(t_1), . . . ,\bm x(t_N)\}$, which serves as the estimate of this quantities that satisfies the model. The dimension of the path is $(N + 1)D$. \cite{abarbanel2022statistical}

The traditional data assimilation method was introduced to estimate parameters and states of a dynamical system\cite{toth2011dynamical} by adding a balanced term to the dynamical equation to synchronize the experimental data and the model, as shown in Eqn \ref{eq:dyn_nudg}. $u_l(t)$ is a time-dependent variable to control the synchronization. Given $u_l(t)>0$, if the estimate $x_l(n)$ is lower than the data $y_l(n)$, the synchronization term $u_l(n)(y_l(n)-x_l(n))$ will be positive and the adjusted dynamical equation will increase $x_l$ conversely, vice versa. Therefore the approach is called ``nudging". At the end of data assimilation, we hope the estimate $x_l(n)$ is synchronized with the data, and the synchronization term nearly vanishes. We can then give predictions without such a term. To meet this condition, a regulation term $u_l(n)^2$ is added to the cost function that prevents the control variable $u_l(n)$ being undesirable large. The cost value now contains two parts: the root-mean-square(RMS) deviation and the new $u_l(n)^2$ term. $R_m$ represents the inverse covariance of measurement error, which can be taken as one if normalization is not required.
\begin{equation}
C(\bm{X},\bm{p}) = \frac{1}{2(N+1)} \{\sum_{n=0}^{N}\sum_{l=1}^L R_m(y_l(n)-x_l(n))^2+u_l(n)^2 \} 
\end{equation}
with the constraints
\begin{equation}\label{eq:dyn_nudg}
x_d(n+1)= f_d(\bm x(n),\bm p) + u_l(n)(y_l(n)-x_l(n)))\delta_{dl}
\end{equation}
Under the data assimilation operation, an optimization process will be performed to search for the minimum cost value. There is a challenge for the traditional method based on value deviation: when applied to experimental data, an imperfect model sometimes makes it impossible to fit spike-timing precisely. As long as the estimated spike does not overlap with the measured spike, the original cost function fails to evaluate the spike-timing correlation because the time-shift distance will not influence the cost value. Therefore we proposed an improved convolutional method to over this weakness.

\subsection{Convolutional Method}
How can we modify the cost function to quantify the time shift of spikes? We make the sharp spikes broader by applying a convolutional kernel to the whole time series so that sightly mismatched spikes will overlap further, and the extent of overlap will quantitatively reflect the spike distance. In this way, spike-timing accuracy can also be evaluated. 
A Gaussian filter with a variance of $\sigma_G^2$ is exerted on the measured data and its estimate. We can approximate the Gaussian filter with the $2\sigma$ range (95\%) in real application for computational efficiency. 
\begin{equation}
\tilde{x}(n)=\sum^N_{n'=0}x(n')g(n-n'),g(n)=\frac{1}{\sqrt{2\pi}\sigma_G}e^{-\frac{n^2}{2\sigma_G^2}}
\end{equation}

Instead of the original data and estimate, we now synchronize the Gaussian-filtered data and estimate. Notice that the state variable inside the dynamical equation $x_d(n+1)= f_d(\bm x(n),\bm p)$ is still the original one since the state variables after convolution does not obey the  dynamics. 
\begin{eqnarray}
C(\bm{X},\bm{p},\sigma_G) = &&\frac{1}{2(N+1)} \{\sum_{n=0}^{N}\sum_{l=1}^L \tilde{R}_m(\tilde{y}_l(n)-\tilde{x}_l(n))^2 \nonumber \\
&&+u_l(n)^2 \} 
\end{eqnarray}
with the constraints
\begin{equation}
x_d(n+1)= f_d(\bm x(n),\bm p) + u_l(n)(\tilde{y}_l(n)-\tilde{x}_l(n)))\delta_{dl}
\end{equation}
The inverse covariance $R_m$ will change after convolution. We define the new one as $\tilde{R}_m$, which value depends on the probability distribution of the data set. The cost value is $\sigma_G$ dependent and acts as a hyperparameter for the optimization. Intuitively, there should be an optimal range of choosing $\sigma_G$: too small would not be sufficient to make the spike width wider, while excessive stretching would undermine the curve pattern.

\subsection{Neural Model}
In 1952, Hodgkin and Huxley developed a prominent dynamical model to describe intracellular neural voltage(potential)\cite{hodgkin1952quantitative}. Neuron was modeled as an RC circuit through which sodium, potassium, and leak currents can pass. Although more iron channels were found, and the related molecular microstructures were proved to vary among species, the Hodgkin-Huxley model remains simple but sufficient to start characterizing electrophysiology behavior. The voltage dynamics are described as:
\begin{eqnarray*}
C_m\frac{dV(t)}{dt} = &&I_{inj}(t) + g_{Na}m(t)^3h(t)(E_{Na}-V(t)) \\
&&+ g_{K}n(t)^4(E_{K}-V(t)) + g_{L}(E_{L}-V(t)) 
\end{eqnarray*}
The three gating variables $a(t) = \{m(t), h(t), n(t)\}, 0 \leqslant a(t) \leqslant 1$ which represents the probability of an individual gating particals being open, follows the 
first order kinetic equation: 
\begin{equation*}
\frac{da(t)}{dt} = \frac{a_{\infty}(V)-a(t)}{\tau_a(V)}
\end{equation*}
The limiting probability $a_{\infty}(V) $ and time constant $\tau_a(V)$ are volatage dependent and satisfies the following equations:
\begin{eqnarray*} 
&a_{\infty}(V) = 0.5(1+tanh(\frac{V-V_a}{\Delta V_a})) \\
&\tau_a(V) = t_{a0}+t_{a1}(1-tanh^2(\frac{V-V_a}{\Delta V_a})) \\
\end{eqnarray*}
$I_{inj}(t)$ comes from the current-clamp experiments\cite{sterratt2011principles}, which allow us to inject a current into the neuron and measure the corresponding voltage response $V(t)$. The injected current should have sufficient amplitudes to generate spikes(typical behavior of the neural system), and its frequency should not exceed the low-pass cutoff frequency since the neuron serves as a low-pass filter(RC circuit). In the data assimilation description, among four state variables $[V(t),m(t),h(t),n(t)]$, only one variable, voltage $V(t)$, is observed. $D=4$, $L=1$. The time-independent parameters differ among species and neuron types, which can be estimated during data assimilation.

\section{Results}
\subsection{Evaluate predictions}
One critical analysis of the data assimilation results is that we want to know if a prediction has high quality or which prediction describes future behavior more precisely. The most common standard is calculating the root-mean-square deviation, which compares the values at each time step and then combines them with a square root. A negligible time
shift between prediction and measurement would be considered inconsistent, thus a poor prediction. This difference will be flattened by replacing the original values with the ones after the Gaussian filter. For such a case common in electrophysiology data, the convolutional method outperforms the traditional cost function in evaluating the performance of different predictions. 

\begin{figure}
\begin{subfigure}{\columnwidth}
\includegraphics[width=\columnwidth]{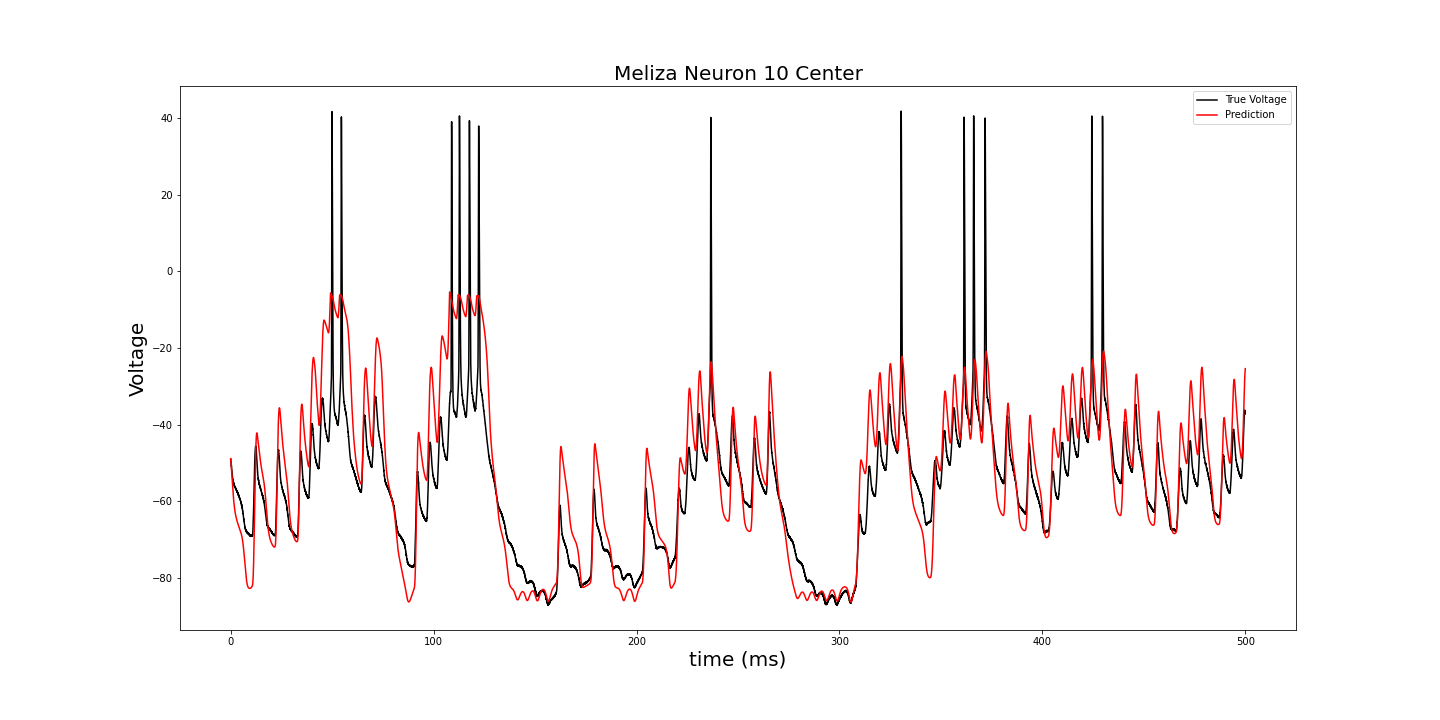}
\caption{Prediction 1}
\end{subfigure}
\begin{subfigure}{\columnwidth}
\includegraphics[width=\columnwidth]{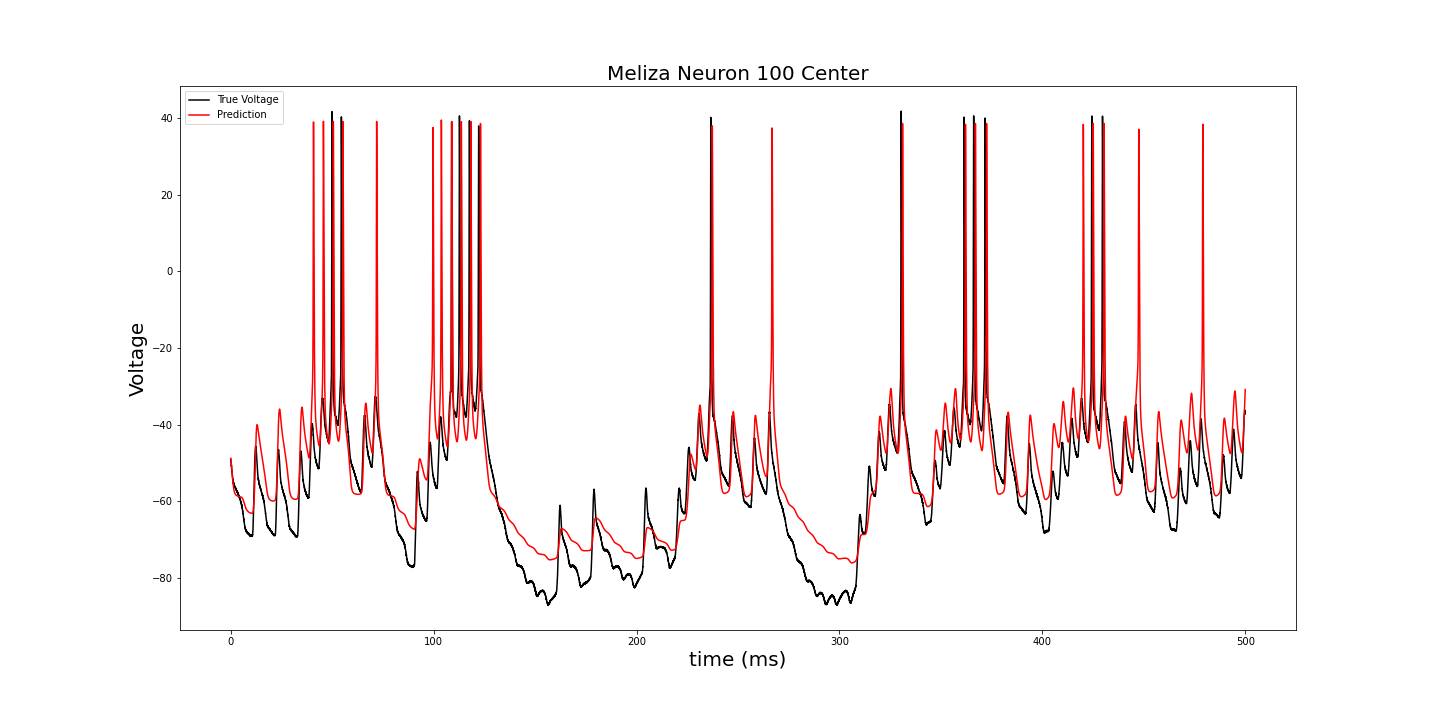}
\caption{Prediction 2}
\end{subfigure}
\caption{\label{fig:pred_comp}Evaluate two predictions(Red) based on the measurements(Black).}
\end{figure}

\begin{figure}
\includegraphics[width=\columnwidth]{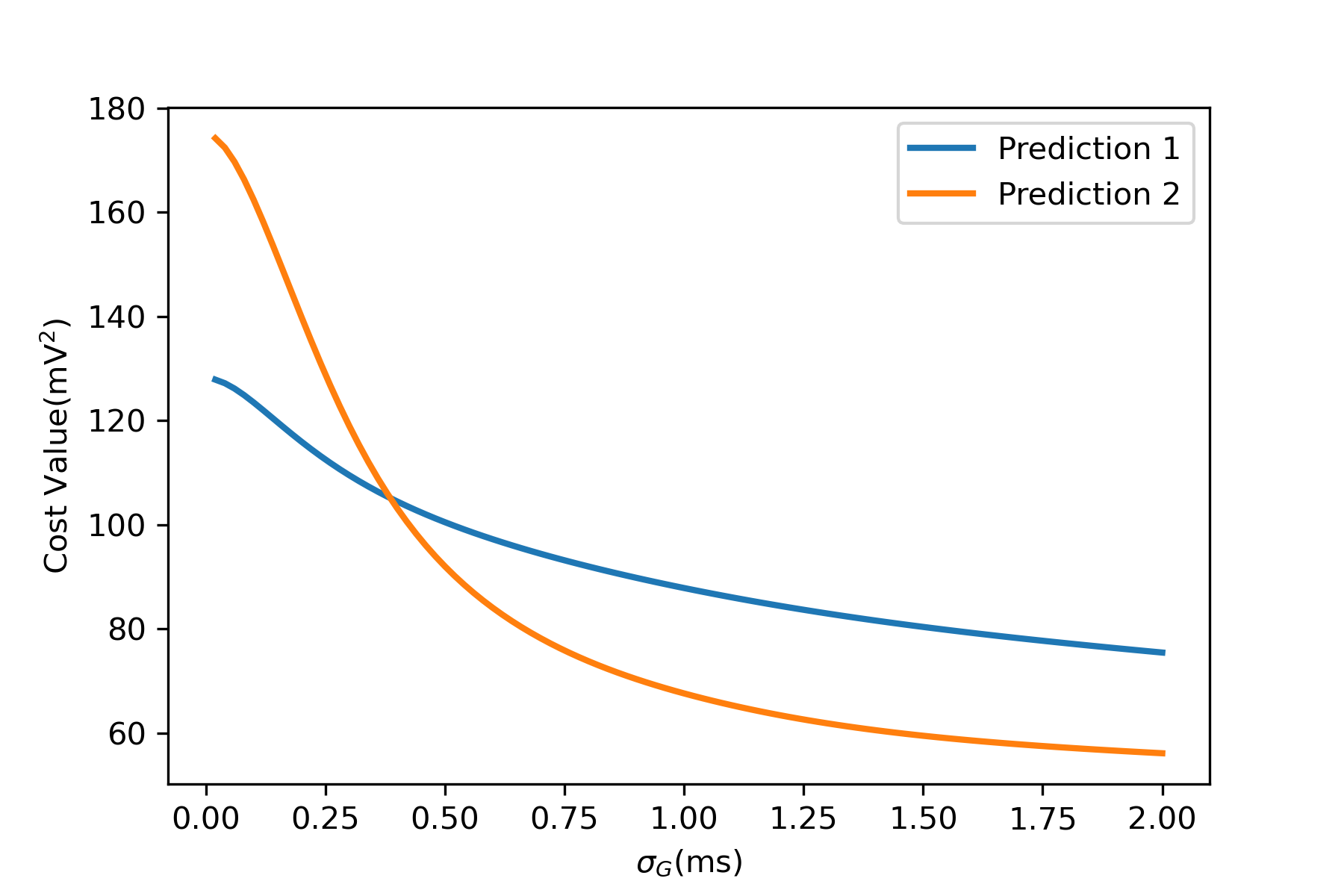}
\caption{\label{fig:cost_sigma}Cost value as the function of $\sigma_G$. Crosspoint shows the cost function of these two methods are paying attention to different features.}
\end{figure}

To illustrate, in Fig \ref{fig:pred_comp}, two predictions are used for comparison. While using the traditional cost function, Prediction 1 has lower cost value than Prediction 2, as shown when $\sigma_G=0ms$ in Fig \ref{fig:cost_sigma}. This unexpected result comes from the truth that Prediction 1 describes the hyperpolarization region more accurately than Prediction 2. Prediction 2, instead, predicts the spikes-timing better. However, the slight spike delay between spikes significantly raises the cost value. As the variance of the Gaussian kernel $\sigma_G$ increases, the cost value will be alternated when $\sigma_G$ approaches 0.5ms or ~1ms order of magnitude. This threshold time recommends an empiric optimal choice of $\sigma_G$ for the convolutional data assimilation method.   

\subsection{Twin experiment}
\begin{figure}
\includegraphics[width=\columnwidth]{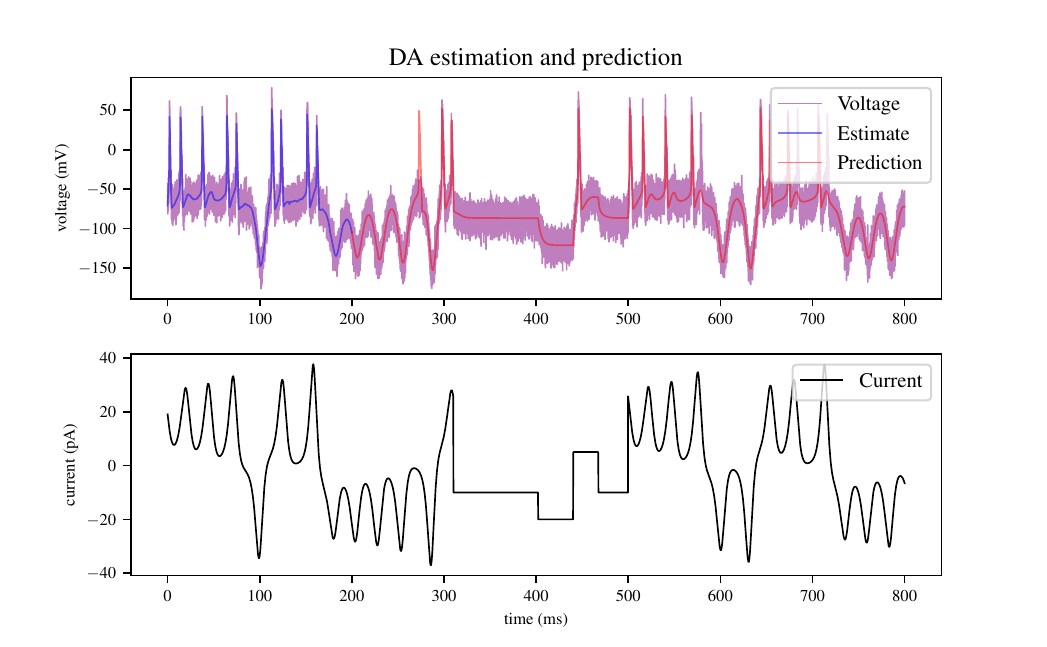}
\caption{\label{fig:twinexp}Twin expriment with 20dB noise added. Convolutional method is applied with $\sigma_G$ = 0.4ms. Purple: Data; Blue: Estimation given by data assimilation; Red: Prediction generated by ODE integrator using estimated parameters.}
\end{figure}

Before applying the convolutional method to experimental data, we first perform a twin experiment, an ``easy task" of data assimilation, where the data is ideally following the dynamical model. In a twin experiment, the voltage data is not measured but instead generated by integrating the dynamical equations forward with selected model parameters and the injected current time series. The known parameters can be used to assess the accuracy of the parameters estimated by the data assimilation method. We use a combination of 1D projection of a Lorentz system and step current as injected current. Gaussian noise with a 20dB signal-noise ratio is added to the generated voltage to simulate the experimental noise and test the robustness of the optimization method. We choose the hyperparameter of the convolutional method, $\sigma_G$, to be 0.4ms. The result is reported in Fig \ref{fig:twinexp} and Tab \ref{tab:twinexp}. The convolutional method is validated to be effective since the prediction fits thoroughly with the generated data, and most of the estimated parameters are comparably close to the true value. Few incompatible parameters make ignorable contributions to the dynamical profile. The success leads us to real experimental data, a ``harder task," demonstrating the convolutional method's benefits.

\begin{table*}%The best place to locate the table environment is directly after its first reference in text
\caption{\label{tab:twinexp}%
Table of parameters. Comparing the estimated parameters and their actual values proves that the convolutional method is effective. The searching regions(lower/upper bounds) where we find these parameter estimates are also provided. 
}
\begin{ruledtabular}
\begin{tabular}{lddddl}
\textrm{Parameters}&
\textrm{Truth}&
\textrm{Estimate}&
\textrm{Lower bound}&
\textrm{Upper bound}&
\textrm{Units}\\
\colrule
$C_m$ & 1.0 & 1.0 & 0.5 & 2.0 & $\mu$F/cm$^2$\\
$g_{Na}$ & 120.0 & 168.2 & 50.0 & 200.0 & mS/cm$^2$\\
$g_K$ & 20.0 & 10.7 & 5.0 & 40.0 & mS/cm$^2$\\
$g_L$ & 0.3 & 0.3 & 0.1 & 1.0 & mS/cm$^2$\\
$E_{Na}$ & 50.0 & 54.0 & 0.0 & 100.0 & mV\\
$E_K$ & -77.0 & -78.6 & -100.0 & -50.0 & mV\\
$E_L$ & -54.4 & -52.4 & -60.0 & -50.0 & mV\\

$V_m$ & -40.0 & -38.8 & -60.0 & -30.0 & mV\\
$\Delta V_m$ & 15.0 & 14.7 & 11.0 & 27.0 & mV\\
$\tau_{m0}$ & 0.10 & 0.23 & 0.05 & 0.25 & ms\\
$\tau_{m1}$ & 0.40 & 0.12 & 0.1 & 1.0 & ms\\

$V_h$ & -55.0 & -57.8 & -70.0 & -40.0 & mV\\
$\Delta V_h$ & 30.0 & 30.8 & 25.0 & 39.0 & mV\\
$\tau_{h0}$ & 1.0 & 0.58 & 0.1 & 5.0 & ms\\
$\tau_{h1}$ & 5.0 & 5.6 & 2.0 & 12.0 & ms\\

$V_n$ & -60.0 & -62.2 & -70.0 & -40.0 & mV\\
$\Delta V_n$ & -15.0 & -17.1 & -20.0 & -8.0 & mV\\
$\tau_{n0}$ & 1.0 & 0.75 & 0.1 & 5.0 & ms\\
$\tau_{n1}$ & 7.0 & 9.4 & 1.0 & 15.0 & ms\\

\end{tabular}
\end{ruledtabular}
\end{table*}

\subsection{Application on Experimental Data}
Experimental data are collected in vitro from a neuron in the brain slice of the CA1 hippocampus region of a healthy mouse in Stutzmann Lab with a 50kHz sampling rate. A designed current flow is injected into the neuron while the corresponding voltage response is recorded simultaneously. This current is a time series that usually contains chaotic and non-chaotic regions to reflect the dynamical system's behavior comprehensively. 

The experimental sampling rate may be too high for our data assimilation purples, adding an unnecessary burden to available computational resources. Remember the path vector $\bm{X}$ has $D(N+1)$ dimension, where $N+1$ is the number of used time steps(data points). Since fewer data points reduce the dimension of the optimization profile and enhance its computational efficiency, we want to determine the optimal extent to which we can downsample the original data with a 50kHz sampling rate. According to the Nyquist-Shannon sampling theorem \cite{shannon1949communication}, the minimum sampling frequency should be two times the highest frequency component of the dataset. The highest frequency component of electrophysiology data is at spikes, excluding noise. The selected time step should be larger than half the spike width to capture the spike activities. Specifically, our electrophysiology data has a spike width of around 3ms. If the original data is downsampled ten times from 50kHz to 5kHz, the 0.2ms step size still satisfies the theorem and spares extra data points to describe the detailed shape of the spikes. An identical downsample preprocessing has been applied when performing data assimilation on electrophysiology data \cite{abarbanel2017unifying}. The overall performance has been verified to be sufficient. For the following results in this section, 5kHz downsampled data are applied. 

\begin{figure*}
\begin{subfigure}{\columnwidth}
\includegraphics[width=\columnwidth]{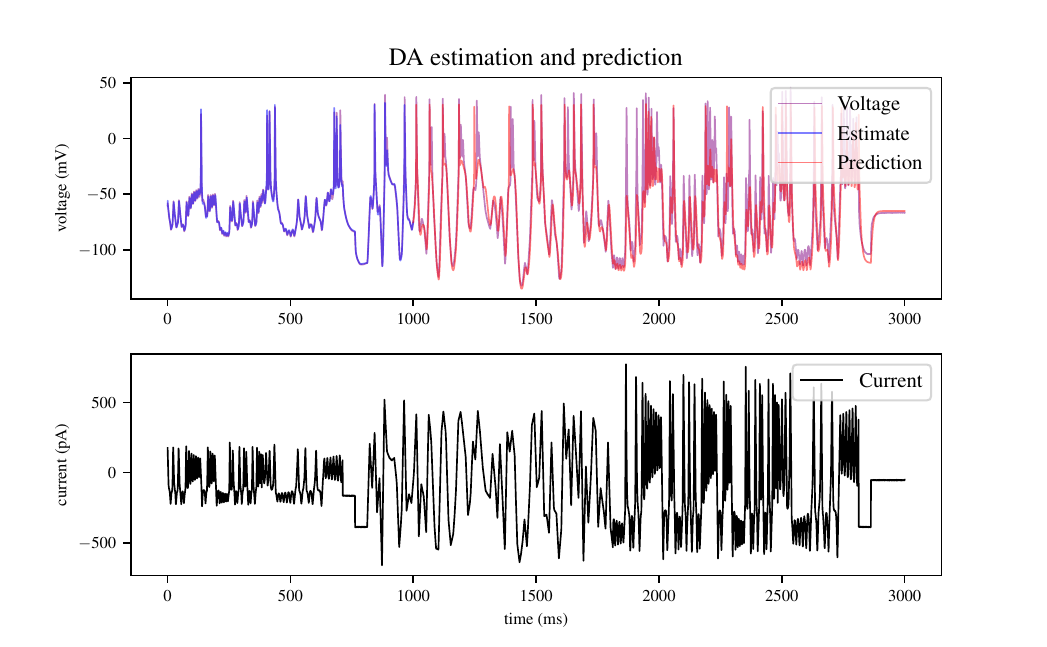}
\caption{\label{fig:comp1_0.8}$\sigma_G$ = 0.8ms}
\end{subfigure}
\begin{subfigure}{\columnwidth}
\includegraphics[width=\columnwidth]{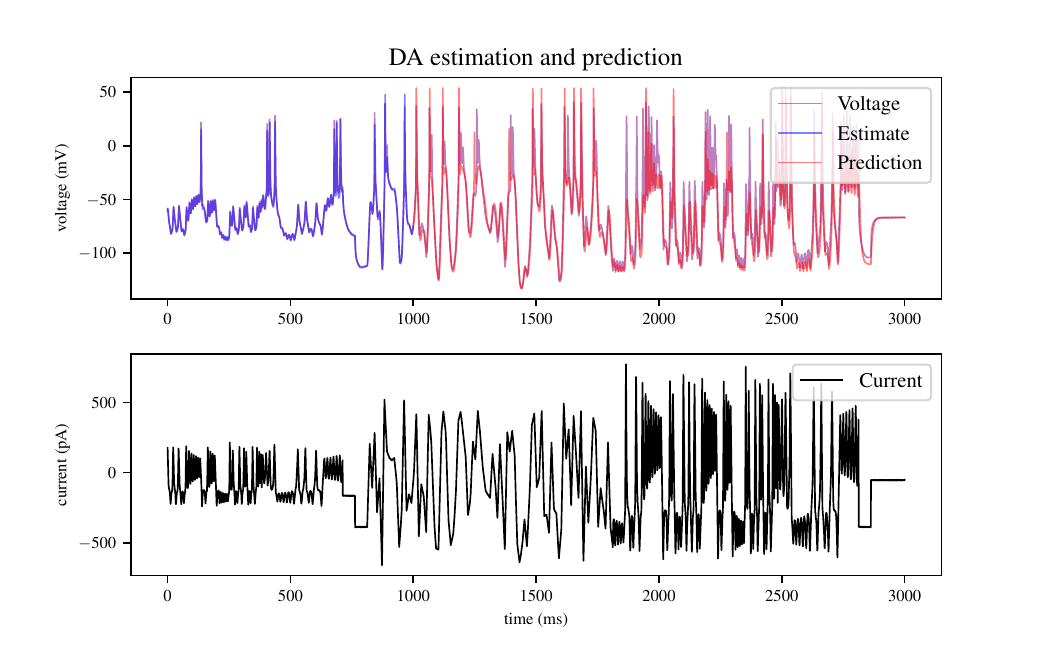}
\caption{\label{fig:comp1_0}No kernel is applied}
\end{subfigure}
\caption{\label{fig:comp1}Purple: Data; Blue: Estimation given by data assimilation; Red: Prediction generated by ODE integrator using estimated parameters. The $\sigma_G$ = 0.8ms convolutional method outperforms the traditional method.}
\end{figure*}

To capture the dynamical features, we select a 1s window with a suitable amount of spikes for estimation. The following 2s window is used to evaluate the prediction performance by integrating the dynamical equations forward using the estimated parameters. Both convolutional(Fig \ref{fig:comp1_0.8}) and traditional(Fig \ref{fig:comp1_0}) methods are implemented for comparison, as shown in Fig \ref{fig:comp1}. Noticeably, The former surpasses the latter in predicting the amplitude of spikes. Although for the experimental data, we do not know the actual values of the parameters, this outperformance indicates that the new method estimates the amplitude-related parameters more faithfully than the previous by smoothing the discrepancy caused by spike mismatch and rebalancing the cost function. The choice of $\sigma_G=$0.8ms is consistent with the previous empirical conclusion.

\begin{figure*}
\begin{subfigure}{\columnwidth}
\includegraphics[width=\columnwidth]{real_c2.pdf}
\caption{\label{fig:comp2_0.8}$\sigma_G$ = 0.8ms}
\end{subfigure}
\begin{subfigure}{\columnwidth}
\includegraphics[width=\columnwidth]{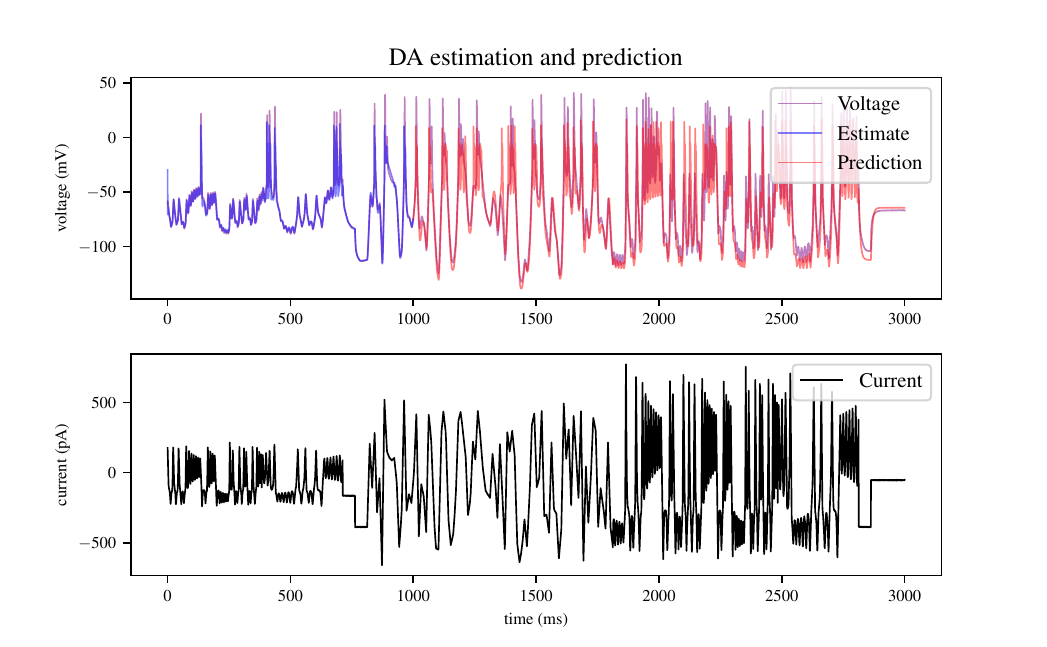}
\caption{\label{fig:comp2_2}$\sigma_G$ = 2ms}
\end{subfigure}
\caption{\label{fig:comp2}Purple: Data; Blue: Estimation given by data assimilation; Red: Prediction generated by ODE integrator using estimated parameters. The $\sigma_G$ = 0.8ms convolutional method outperforms The $\sigma_G$ = 2ms one.}
\end{figure*}

However, if we continuously increase $\sigma_G$, this improvement will not last forever. Fig \ref{fig:comp2_2}(compared with Fig \ref{fig:comp2_0.8}) shows that if the $\sigma_G$ we pick is too large, the excessive extent that spikes stretch out will instead reduce the performance quality. An extreme example is if we choose $\sigma_G$ to be infinity, all the dynamical behavior will be removed and replaced by a step function of average value. Choose a wise $\sigma_G$ will half the effort.

% \begin{figure}
% \includegraphics[width=\columnwidth]{real_c2.pdf}
% \caption{\label{fig:real0.8ms}$\sigma_G$ = 0.8ms. Prediction generated by ODE integrator using estimated parameters.}
% \end{figure}

% \begin{figure}
% \includegraphics[width=\columnwidth]{real_c0.pdf}
% \caption{\label{fig:real}No kernel is applied. Prediction generated by ODE integrator using estimated parameters.}
% \end{figure}

% \begin{figure}
% \includegraphics[width=\columnwidth]{real_c5.pdf}
% \caption{\label{fig:real2ms}$\sigma_G$ = 2ms. Prediction generated by ODE integrator using estimated parameters.}
% \end{figure}

\section{Discussion}
\subsection{Optimal $\sigma_G$}
Given the importance of selecting $\sigma_G$, we come up with the question: Can we design an optimization method that can automatically determine the optimal value of this hyperparameter? Ideally, if the cost function is well normalized with $\tilde{R}_m$, there should be optimal $\sigma_G$ where we can find the minimum cost value. 

Now we need to apply the optimization twice. For each $\sigma_G$, we use the convolutional method to find the minimum of cost function. Using Monte Carlo method, we can start with a initial ${\sigma_G}_0$ and iteratively generates solutions by making perturbations to $\sigma_G$, such as adding a normally distributed random number with zero mean and standard deviation $\delta$, to $\sigma_G$(Notice $\sigma_G$ can not be negative). The perturbation size $\delta$ can be adjusted based on the desired exploration and exploitation trade-off. After generating a new solution, the Metropolis-Hastings acceptance standard is used to determine whether to accept or reject the new solution. The Metropolis-Hastings acceptance standard takes into account the difference in objective function values between the current and new solutions, as well as a temperature parameter that controls the acceptance probability. As the algorithm progresses, the temperature parameter is decreased to make the algorithm more selective in accepting new solutions and converge to a global minimum.

By iteratively generating and accepting or rejecting new solutions, the Monte Carlo method can explore the search space and fine-tune the value of $\sigma_G$ to approach the optimal value for the given optimization problem. We can also find the minimum of cost function with the optimal $\sigma_G$.

However, normalize the $\tilde{R}_m$ is challenging since the probability distribution is not typical. We will retain this idea for future work.

\subsection{Application outside of neuroscience}
The method can be used to synchronize all signals that have sharp patterns with temporal structure. These signals are typically associated with the activity of membrane potentials, but they can also be observed in other types of signals that exhibit a similar discrete, binary structure. Some examples include:
\begin{itemize}
    \item Action potentials in muscle fibers: Like neurons, muscle fibers can generate action potentials, which are brief, all-or-nothing electrical events that result in a contraction of the muscle. These action potentials can be recorded using electromyography (EMG) electrodes.
    \item Event-related potentials (ERPs) in electroencephalography (EEG): ERPs are brief electrical responses in the brain that occur in response to a specific sensory, cognitive, or motor event. They are often recorded using EEG electrodes and each spike corresponding to the occurrence of an ERP.
    \item Glucose fluctuations in diabetes: In people with diabetes, glucose levels can fluctuate rapidly and unpredictably. These fluctuations can be detected using continuous glucose monitoring (CGM) devices and each spike corresponding to a glucose measurement.
    \item Geophysics: Seismographs measure seismic waves generated by earthquakes and other events. The resulting signals can exhibit spikes or bursts of activity, which can be analyzed to understand the properties of the seismic event.
    \item Speech processing: When analyzing speech signals, it is common to represent the signal as a series of discrete events called phonemes. The timing of these phonemes can be represented as a spike train, which can be used to study speech perception and production.
    \item Astronomy: When observing astronomical events such as supernovae or gamma-ray bursts, the resulting signals can exhibit sharp, transient bursts of activity that can be analyzed to understand the properties of the event.
    \item Finance: In financial markets, the timing of trades can be represented as a spike train, with each spike corresponding to a trade. This representation can be used to analyze market activity and study the behavior of traders.
    \item Ecology: When studying the behavior of animals, researchers may use radio collars or other tracking devices to record the animal's movement. The resulting signals can exhibit spikes or bursts of activity, which can be analyzed to understand the animal's behavior and ecology.
    
\end{itemize}

\newpage

\bibliography{minAc}% Produces the bibliography via BibTeX.

\end{document}